\documentclass{INTERSPEECH2023}
\usepackage{multirow}
\usepackage{booktabs}
\usepackage{comment}

\interspeechcameraready

\title{Unsupervised speech enhancement with deep dynamical \\ generative speech and noise models}
\name{Xiaoyu Lin$^1$, Simon Leglaive$^2$, Laurent Girin$^3$, Xavier Alameda-Pineda$^1$ \thanks{This research was supported by ANR-3IA MIAI (ANR-19-P3IA-0003), ANR-JCJC ML3RI (ANR-19-CE33-0008-01), H2020 SPRING (funded by EC under GA \#871245).}}
\address{$^1$ Inria Grenoble Rh\^one-Alpes, Univ. Grenoble Alpes, France \\
$^2$ CentraleSupélec, IETR (UMR CNRS 6164), France\\
$^3$ Univ. Grenoble Alpes, CNRS, Grenoble-INP, GIPSA-lab, France
}
\email{}

\begin{document}

\maketitle
 
\begin{abstract}
This work builds on a previous work on unsupervised speech enhancement using a dynamical variational autoencoder (DVAE) as the clean speech model and non-negative matrix factorization (NMF) as the noise model. We propose to replace the NMF noise model with a deep dynamical generative model (DDGM) depending either on the DVAE latent variables, or on the noisy observations, or on both. 
This DDGM can be trained in three configurations: noise-agnostic, noise-dependent and noise adaptation after noise-dependent training. 
Experimental results show that the proposed method achieves competitive performance compared to state-of-the-art unsupervised speech enhancement methods, while the noise-dependent training configuration yields a much more time-efficient inference process.
\end{abstract}

\section{Introduction}

Speech enhancement is a fundamental task of speech processing that aims at recovering the clean speech signal from a noisy audio recording \cite{speechenhancement, loizou2007speech}. In recent years, methods based on deep neural networks 
(DNNs) have greatly advanced research in this field. The most widely-used approach is a direct \textit{supervised} mapping from the noisy speech signal to either the clean speech target or a denoising mask, see a review in \cite{8369155}.
While this approach has shown impressive results, it also has limitations: it requires a huge amount of paired noisy-clean speech data for training and can show poor generalization to noise types and acoustic conditions unseen during training. Another type of supervised method resort to generative adversarial networks (GANs)
to learn a conditional distribution of the clean speech signal given the noisy speech \cite{Pascual2017SEGANSE,fu2019metricGAN, Fu2021MetricGANAI}. Recently, methods based on diffusion models were proposed in, e.g., \cite{9746901, richter2022speech}. 
A diffusion model transforms a clean signal into a noisy one by adding noise step by step, and speech enhancement is obtained by applying the inverse diffusion process conditioned on the input noisy speech signal. These methods show good generalization capability, but still require a large amount of paired data for training and are quite slow at inference.

\textit{Unsupervised} speech enhancement methods were recently developed to improve the performance of models on unseen noise types. Here, the models do not use parallel clean-noisy data for training. Instead, they use either non-parallel noisy-clean data (i.e., the clean and noisy samples do not correspond) \cite{9104020, 9689669}, or clean data only \cite{8461530, 8516711, 9894060}, or noisy data only \cite{Alamdari2019ImprovingDS, Kashyap2021SpeechDW, 9616166, 9747180}. Unsupervised  speech enhancement methods can be further divided into \textit{noise-dependent} (ND) and \textit{noise-agnostic} (NA) methods~\cite{9894060}. ND methods use noise or noisy speech training samples to learn some noise characteristics. In contrast, NA methods only use clean speech signals for training and the noise characteristics are estimated at test time  for each noisy speech sequence to process. A typical unsupervised NA approach uses a pre-trained variational autoencoder (VAE) as a prior distribution of the clean speech signal and a non-negative matrix factorization (NMF) model for the noise variance \cite{8461530, 8516711, Pariente2019ASP, 10095237}. The NMF parameters and the VAE latent vector are estimated at test time from the noisy signal and combined to build a denoising Wiener filter. Further developments in this general line were proposed in, e.g., \cite{8682546, 9414363, 8683704, 9747036}.

Recently, \cite{9894060} proposed to replace the VAE by a dynamical variational autoencoder (DVAE) \cite{MAL-089, Bie2021ABO}, yielding better clean speech modeling by considering the temporal dependencies across successive spectrogram frames. The algorithm proposed in \cite{9894060}  was shown to achieve very competitive performance even when compared to supervised approaches. However, this algorithm has two main drawbacks. First, the NMF may be a too simple model for many real-world noise signals, which are poorly described in the spectrogram domain as a non-negative linear combination of a few spectral templates. Second, at test time, the inference algorithm, which must be run on each noisy sequence independently, is very time-consuming.

In this work, we aim at both increasing the modeling power of the noise model and accelerating the inference process. To this aim, we build on \cite{9894060} and propose to replace the NMF noise model with a deep dynamical generative model (DDGM), which is a general class of dynamical models for the generation of sequential data based on DNNs.\footnote{Note that the DVAE used for clean speech modeling also belongs to the DDGM family, hence the paper title.} We implement and test the DDGM noise model with dependencies either on: the DVAE latent variables (LV), or the noisy observations (NO), or both (NOLV). 
Moreover, these three variants are implemented and tested in both ND (using a large noisy speech dataset) or NA configurations. 
Even further, the models trained in ND configuration can then be fine-tuned on each noisy speech test sequence to get adapted to specific noise types in the test set (i.e., ND followed by noise adaptation).
Experimental results show that the proposed method obtains performance that is comparable to that of \cite{9894060},
while in the ND configuration, it requires much less computation time during inference. 

\begin{figure*}[htb!]
    \centering
    \includegraphics[width=\linewidth]{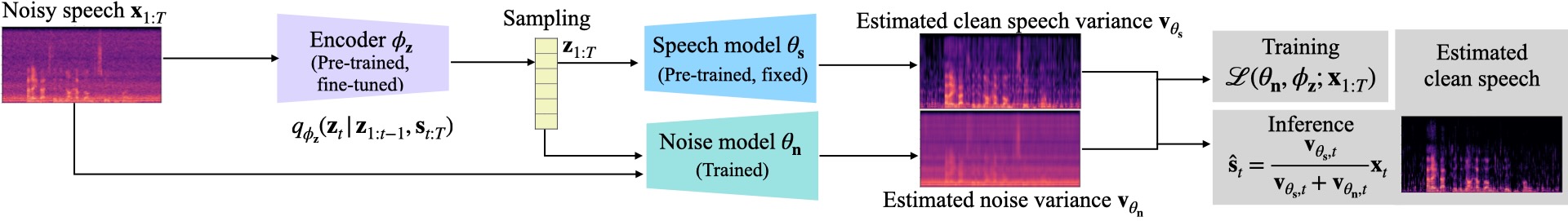}
    \caption{Schematic view of the proposed speech enhancement method.}
    \label{fig:model_architecture}
\end{figure*}

\section{Proposed method}

In this section, we present our unsupervised speech enhancement method using DDGM-based speech and noise models.

\subsection{Clean speech modeling with an RVAE}
\label{subsec:clean-RVAE}

We work in the short-time Fourier transform (STFT) domain. Let $\mathbf{s}_{1:T}=\{\mathbf{s}_t\}^T_{t=1} \in \mathbb{C}^{F \times T}$ denote the STFT spectrogram of the clean speech. Each vector $\mathbf{s}_t=\{s_{t,f}\}^F_{f=1} \in \mathbb{C}^{F}$ is the short-time spectrum at time frame $t$, and $f$ denotes the frequency bin. Let $\mathbf{z}_{1:T} \in \mathbb{R}^{L \times T}$ denote the associated latent vector sequence, with $L \ll F$ the latent dimension. As reported in \cite{9894060}, among several tested DVAE models, the recurrent variational autoencoder (RVAE) model \cite{9053164} worked best on the speech enhancement task. So, we also use this model in this work. The RVAE generative model is defined as:
\begin{equation}\label{eq:generative_rvae}
    p_{\theta_{\mathbf{s}}}(\mathbf{s}_{1:T}, \mathbf{z}_{1:T}) = \prod_{t=1}^T p_{\theta_{\mathbf{s}}}(\mathbf{s}_t|\mathbf{z}_{1:t})p(\mathbf{z}_t).
\end{equation}
For each time frame $t$, and conditionally to $\mathbf{z}_{1:t}$, $\mathbf{s}_t$ is assumed to follow a circularly-symmetric zero-mean complex Gaussian distribution \cite{6797100, 5720325},
$p_{\theta_{\mathbf{s}}}(\mathbf{s}_t|\mathbf{z}_{1:t}) = \mathcal{N}_c\big(\mathbf{s}_t; \mathbf{0}, \textrm{diag} (\mathbf{v}_{\theta_{\mathbf{s}},t})\big).$
In this work, all covariance matrices are assumed diagonal and are represented by the vector of diagonal entries. Here, $\mathbf{v}_{\theta_{\mathbf{s}},t} \in \mathbb{R}_+^{F}$ is a function of $\mathbf{z}_{1:t}$ and is modeled with the RVAE decoder. The latent vector $\mathbf{z}_t$ is assumed to follow a standard Gaussian prior distribution, $p(\mathbf{z}_t) = \mathcal{N}(\mathbf{z}_t; \mathbf{0}, \mathbf{I}).$

The inference model, i.e.~the approximated posterior distribution of RVAE, is defined as \cite{9053164}:
\begin{equation}
    q_{\phi_{\mathbf{z}}} (\mathbf{z}_{1:T}|\mathbf{s}_{1:T}) = \prod_{t=1}^T q_{\phi_{\mathbf{z}}} (\mathbf{z}_t|\mathbf{z}_{1:t-1}, \mathbf{s}_{t:T}),
\end{equation}
with $\mathbf{z}_t$ assumed to follow a (real-valued) Gaussian distribution, $q_{\phi_{\mathbf{z}}} (\mathbf{z}_t|\mathbf{z}_{1:t-1}, \mathbf{s}_{t:T}) = \mathcal{N} \big(\mathbf{z}_t; \boldsymbol{\mu}_{\phi_{\mathbf{z}},t}, \textrm{diag} (\mathbf{v}_{\phi_{\mathbf{z}},t})\big)$,
where $\boldsymbol{\mu}_{\phi_{\mathbf{z}},t} \in \mathbb{R}^L$ and $\mathbf{v}_{\phi_{\mathbf{z}},t} \in \mathbb{R}_+^L$ are both a function of $\mathbf{z}_{1:t-1}$ and $\mathbf{s}_{t:T}$, which is modeled with the RVAE encoder.\footnote{In practice, the squared modulus of the $\mathbf{s}_{t:T}$ entries are send to the encoder input instead of the complex-valued STFT coefficients.}  

The RVAE model is pre-trained on a clean speech dataset by maximizing the evidence lower bound (ELBO) \cite{9053164}:
\begin{align} \label{eq:RVAE-ELBO}
    \mathcal{L} (\theta_{\mathbf{s}}, \phi_{\mathbf{z}}; \mathbf{s}_{1:T}) 
    &= - \sum^T_{t=1} \mathbb{E}_{q_{\phi_{\mathbf{z}}}} \Big [ d_{\textrm{IS}} (|\mathbf{s}_t|^2, \mathbf{v}_{\theta_{\mathbf{s}}, t}) \nonumber \\
    & + D_{\textrm{KL}} \big( q_{\phi_{\mathbf{z}}} (\mathbf{z}_t|\mathbf{z}_{1:t-1}, \mathbf{s}_{t:T}) || p(\mathbf{z}_t) \big) \Big],
\end{align}
where modulus and exponentiation are element-wise, $d_{\text{IS}}(\cdot,\cdot) $ is the Itakura-Saito divergence \cite{6797100} and $D_{\text{KL}}(\cdot||\cdot)$ is the Kullback–Leibler divergence (KLD).

\subsection{DDGM-based noise model}
Let $\mathbf{x}_{1:T} = \{\mathbf{x}_t\}^T_{t=1} \in \mathbb{C}^{F \times T}$ and $\mathbf{n}_{1:T} = \{\mathbf{n}_t\}^T_{t=1} \in \mathbb{C}^{F \times T}$ denote respectively the complex-valued STFT spectrogram of the noisy speech and the noise, which is assumed additive:
\begin{equation}\label{eq:additive_noise}
    \mathbf{x}_{1:T} = \mathbf{s}_{1:T} + \mathbf{n}_{1:T}.
\end{equation}
At each time frame $t$, $\mathbf{n}_t$ is assumed to follow a circularly-symmetric zero-mean complex Gaussian distribution:
\begin{equation}\label{eq:noise_model}
    p_{\theta_{\mathbf{n}}}(\mathbf{n}_t) = \mathcal{N}_c \big(\mathbf{n}_t; \mathbf{0}, \textrm{diag} (\mathbf{v}_{\theta_{\mathbf{n}},t})\big),
\end{equation}
where $\mathbf{v}_{\theta_{\mathbf{n}},t} \in \mathbb{R}_+^F$ is the noise variance vector. In several previous works, $\mathbf{v}_{\theta_{\mathbf{n}},t}$ was modeled with NMF, i.e.~factorized into the product of two low-rank non-negative matrices. In this work, we model $\mathbf{v}_{\theta_{\mathbf{n}},t}$ with a DDGM. We propose three different noise model dependencies: (i) \textbf{DVAE latent variables~(LV)}, in which $\mathbf{v}_{\theta_{\mathbf{n}},t}$ is a function of the whole sequence of the DVAE latent vectors, i.e.~$\mathbf{v}_{\theta_{\mathbf{n}},t}=\mathbf{v}_{\theta_{\mathbf{n}},t} (\mathbf{z}_{1:T})$;
(ii) \textbf{noisy observations~(NO)}, in which $\mathbf{v}_{\theta_{\mathbf{n}},t}$ is a function of all the past values of the noisy speech, i.e.~$\mathbf{v}_{\theta_{\mathbf{n}},t}=\mathbf{v}_{\theta_{\mathbf{n}},t} (\mathbf{x}_{1:t-1})$; 
and (iii) \textbf{both noisy observations and DVAE latent variables~(NOLV)}, in which $\mathbf{v}_{\theta_{\mathbf{n}},t}$ is a function of all the past values of the noisy speech as well as the past and present values of the DVAE latent vectors, i.e.~$\mathbf{v}_{\theta_{\mathbf{n}},t}=\mathbf{v}_{\theta_{\mathbf{n}},t} (\mathbf{x}_{1:t-1}, \mathbf{z}_{1:t})$. 
For clarity of presentation, let $\mathbf{p}_t$ denote the input of the noise model, i.e.~$\mathbf{p}_t = \mathbf{z}_{1:T}$ in LV, $\mathbf{p}_t = \mathbf{x}_{1:t-1}$ in NO, and $\mathbf{p}_t = \{\mathbf{x}_{1:t-1}, \mathbf{z}_{1:t}\}$ in NOLV.
For all model dependencies (NO, LV, or NOLV), the noise variance $\mathbf{v}_{\theta_{\mathbf{n}},t}$ is a function of $\mathbf{p}_t$ that is implemented by a DNN.

Applying the chain rule and taking into account the conditional dependencies, the generative model over the set of variables  $\{\mathbf{x}_{1:T},\mathbf{s}_{1:T},\mathbf{z}_{1:T}\}$ is given by:
\begin{equation}
\label{eq:gen-model-xsz}
    p_{\theta}(\mathbf{x}_{1:T}, \mathbf{s}_{1:T}, \mathbf{z}_{1:T}) = \prod^T_{t=1} p_{\theta_{\mathbf{n}}}(\mathbf{x}_t | \mathbf{s}_t, \mathbf{p}_t)p_{\theta_{\mathbf{s}}}(\mathbf{s}_t|\mathbf{z}_{1:t})p(\mathbf{z}_t),
\end{equation}
where
\begin{equation} \label{eq:generative-mixture}
    p_{\theta_{\mathbf{n}}}(\mathbf{x}_t|\mathbf{s}_t, \mathbf{p}_t) = \mathcal{N}_c \big(\mathbf{x}_t; \mathbf{s}_t, \textrm{diag} (\mathbf{v}_{\theta_{\mathbf{n}},t} (\mathbf{p}_t))\big)
\end{equation}
is deduced from \eqref{eq:additive_noise} and \eqref{eq:noise_model}, $p_{\theta_{\mathbf{s}}}(\mathbf{s}_t|\mathbf{z}_{1:t})$ and $p(\mathbf{z}_t)$ are  defined in Section~\ref{subsec:clean-RVAE}, and $\theta = \theta_{\mathbf{s}} \cup \theta_{\mathbf{n}}$. 

\subsection{Speech enhancement with the inference model}
The posterior distribution corresponding to the generative model \eqref{eq:gen-model-xsz} factorizes as follows:
\begin{equation}
 \label{eq:inf-mod-sz}   p_\theta(\mathbf{s}_{1:T}, \mathbf{z}_{1:T} | \mathbf{x}_{1:T}) = \prod^T_{t=1} p_{\theta}(\mathbf{s}_t|\mathbf{z}_{1:t}, \mathbf{x}_{t}, \mathbf{p}_t)p_{\theta}(\mathbf{z}_t|\mathbf{z}_{1:t-1}, \mathbf{x}_{1:T}).
\end{equation}
For each time frame $t$, $p_{\theta}(\mathbf{s}_t|\mathbf{z}_{1:t}, \mathbf{x}_{t}, \mathbf{p}_t)$ can be computed in closed form as a complex Gaussian distribution $p_{\theta}(\mathbf{s}_t|\mathbf{z}_{1:t}, \mathbf{x}_t, \mathbf{p}_t) = \mathcal{N}_c(\mathbf{s}_t; \boldsymbol{\mu}_{\theta, t},  \textrm{diag} (\mathbf{v}_{\theta, t}))$, with
\begin{equation} \label{eq:mu_phi_s}
    \boldsymbol{\mu}_{\theta, t} = \frac{\mathbf{v}_{\theta_{\mathbf{s}}, t} ( \mathbf{z}_{1:t})}{\mathbf{v}_{\theta_{\mathbf{s}}, t} ( \mathbf{z}_{1:t}) + \mathbf{v}_{\theta_{\mathbf{n}}, t}(\mathbf{p}_{t}) } \mathbf{x}_t,
\end{equation}
\begin{equation} \label{eq:v_phi_s}
    \mathbf{v}_{\theta,t} = \frac{\mathbf{v}_{\theta_{\mathbf{s}}, t} ( \mathbf{z}_{1:t})\mathbf{v}_{\theta_{\mathbf{n}}, t}(\mathbf{p}_{t})}{\mathbf{v}_{\theta_{\mathbf{s}}, t} (\mathbf{z}_{1:t}) + \mathbf{v}_{\theta_{\mathbf{n}}, t}(\mathbf{p}_{t})},
\end{equation}
where vector multiplication and division are element-wise. Eq.~\eqref{eq:mu_phi_s} provides the clean speech signal minimum mean squared error (MMSE) estimate, which corresponds to the Wiener filter output.
The distribution $p_{\theta}(\mathbf{z}_t|\mathbf{z}_{1:t-1}, \mathbf{x}_{1:T})$ is intractable and cannot be used directly to recursively provide the $\mathbf{z}_{1:T}$ estimate. We thus approximate it with the RVAE inference model (defined in Section~\ref{subsec:clean-RVAE}):
\begin{equation}
    p_{\theta}(\mathbf{z}_t|\mathbf{z}_{1:t-1}, \mathbf{x}_{1:T}) \approx q_{\phi_{\mathbf{z}}}(\mathbf{z}_t|\mathbf{z}_{1:t-1}, \mathbf{x}_{t:T}).
    \label{eq:post-approx}
\end{equation}
Here, the RVAE encoder, pre-trained on a clean speech signal dataset, takes as input the noisy speech signal, and must thus be adapted to such kind of input (see the next sub-section). In the following, we inject \eqref{eq:post-approx} into~(\ref{eq:inf-mod-sz}), and the resulting approximate joint posterior is denoted by $p_{\theta,\phi_{\mathbf{z}}}(\mathbf{s}_{1:T}, \mathbf{z}_{1:T} | \mathbf{x}_{1:T})$. 


\subsection{Model optimization}
\label{subsec:model-optim}

We recall that the parameters $\{\theta_{\mathbf{s}},\phi_{\mathbf{z}}\}$ are learned by pre-training the RVAE on a clean speech dataset. $\theta_{\mathbf{s}}$ is then fixed during the speech enhancement stage, whereas $\phi_{\mathbf{z}}$ has to be fine-tuned on the noisy signal(s), and we also have to estimate the noise model parameters $ \theta_{\mathbf{n}}$. As is usually done in variational inference algorithms, the parameters are optimized by maximizing the ELBO, which is here defined as:
\begin{equation}
\hspace{-3mm} \mathcal{L}(\theta_{\mathbf{n}}, \phi_{\mathbf{z}}; \mathbf{x}_{1:T}) 
    = \mathbb{E}_{p_{\theta, \phi_{\mathbf{z}}}}
    \Big[ \log \frac{p_{\theta}(\mathbf{x}_{1:T}, \mathbf{s}_{1:T}, \mathbf{z}_{1:T})}
    {p_{\theta, \phi_{\mathbf{z}}}(\mathbf{s}_{1:T}, \mathbf{z}_{1:T} | \mathbf{x}_{1:T})} 
    \Big].\hspace{-3mm}
    \label{eq:elbo-inference}
\end{equation}
Given the factorizations \eqref{eq:gen-model-xsz} and \eqref{eq:inf-mod-sz}, and the fact that all involved distributions are Gaussian, \eqref{eq:elbo-inference} can be developed as:
\begin{align}
    \mathcal{L}(\theta_{\mathbf{n}}, \phi_{\mathbf{z}}; \mathbf{x}_{1:T}) &= - \sum_{t=1}^T \mathbb{E}_{q_{\phi_{\mathbf{z}}}} \Big[ 
    d_{\textrm{IS}} (|\mathbf{x}_t|^2, \mathbf{v}_{\theta_{\mathbf{s}},t} + \mathbf{v}_{\theta_{\mathbf{n}},t}) \nonumber \\
    &+ D_{\textrm{KL}} \big(q_{\phi_{\mathbf{z}}}(\mathbf{z}_t|\mathbf{z}_{1:t-1}, \mathbf{x}_{t:T}) || p(\mathbf{z}_t)\big) 
    \Big].\label{ELBO-developed}
\end{align} 

As mentioned before, the model can be trained in either NA or ND configuration. When trained in NA configuration, the parameters $\{\theta_{\mathbf{n}}, \phi_{\mathbf{z}}\}$ are estimated directly from the noisy speech sequence to be enhanced. This is done by optimizing the ELBO \eqref{ELBO-developed} independently on each single noisy speech sequence for a certain number of iterations. Afterwards, the clean speech estimate is computed with \eqref{eq:mu_phi_s}, using the optimal parameters and latent vectors sampled from the encoder. This configuration allows the model to adapt to the specific noise patterns of each test sequence, without the need for any prior knowledge or training data on the noise type. This makes it suitable for scenarios where the noise type is unknown. When trained in ND configuration, the model parameters are estimated by optimizing the ELBO \eqref{ELBO-developed} on a large noisy speech training set using stochastic gradient descent (SGD) optimization (we recall that no parallel noisy-clean data is thus used). Then at test time, the clean speech is computed using \eqref{eq:mu_phi_s} with a single forward pass of the model on the noisy test sequence. This results in a much more time-efficient inference than methods based on an NMF noise model, while still achieving competitive performance. A schematic view of the proposed method is shown in Fig.~\ref{fig:model_architecture}.

\section{Experiments}
\subsection{Datasets and pre-processing}
We used two datasets to evaluate the proposed method: the WSJ0-QUT dataset introduced in  \cite{9053164} and reused in \cite{9894060}, and the publicly available VoiceBank-DEMAND (VB-DMD) dataset \cite{Valentini-Botinhao+2016}. WSJ0-QUT is obtained by mixing clean signals from the Wall Street Journal (WSJ0) dataset \cite{WSJ0} with various types of noise signals from the QUT-NOISE dataset \cite{Dean2015TheQP} with three different signal-to-noise ratio (SNR) values: $-$5, 0 and 5~dB. It contains 12,765 utterances from 101 speakers, 1,026 utterances from 10 speakers and 651 utterances from 8 speakers for model training, validation and test, respectively. VB-DMD is obtained by mixing clean signals from the VoiceBank (VB) corpus \cite{Veaux2013TheVB} with ten types of noise from the DEMAND noise dataset \cite{Thiemann2013TheDE}. Following \cite{9747180}, we used
10,802 utterances from 26 speakers for training, 770 utterances from 2 other speakers for validation, and 824 utterances from 2 other speakers for test. 
The SNR values used for the training set are 15, 10, 5 and 0 dB, while the SNR values used for the test set are 17.5, 12.5, 7.5, and 2.5 dB. 
For each dataset, we first pre-trained the RVAE model on the clean speech dataset, i.e.~WSJ0 or VB; then we estimated the noise model parameters using the noisy speech data, either in the NA or ND configuration (see Section~\ref{subsec:model-optim}
).

Before being input into the neural networks, the audio signals are pre-processed as follows. We compute the STFT with a 64-ms sine window (1,024 samples) and a 75\%-overlap (256-sample shift), resulting in a sequence of 513-dimensional discrete Fourier coefficients (for positive frequencies). The squared modulus of the STFT coefficients is computed afterwards. For the RVAE pre-training and the speech enhancement model trained in ND configuration, we first use a voice activity detection threshold of 30 dB to remove silence portions at the beginning and the end of the signals, and rescale the waveforms in $[-1, 1]$ before computing the STFT coefficients. And we also split the training utterances into smaller sequences of length $T=100$ frames. At test time, the model is evaluated on the complete noisy test utterances, which can be of variable length. The speech enhancement model in NA configuration is trained and evaluated directly on each single complete noisy test utterances.

\begin{table*}[ht!]
\caption{Speech enhancement results. S stands for supervised, U-NA stands for unsupervised noise-agnostic and U-ND stands for unsupervised noise-dependent, U-NDA stands for U-ND training followed by noise adaptation fine-tuning. Except for the RTF, the baselines scores are taken from the corresponding papers. The best scores are in bold and the second best scores are underlined.} 
\centering
\resizebox{\linewidth}{!}{
 \begin{tabular}{ccccccccccc} 
\toprule[1pt]
Data & Model & Sup & SI-SDR $\uparrow$ & PESQ$_{\text{MOS}}$  $\uparrow$ & PESQ$_{\text{WB}}$ $\uparrow$ & PESQ$_{\text{NB}}$ $\uparrow$ & ESTOI $\uparrow$ & \# Iter. $\downarrow$ & RTF $\downarrow$ \\
\midrule
\multirow{7}*{\rotatebox{90}{WSJ0-QUT}} & Noisy mix. & - & -2.6 & 1.83 & 1.14 & 1.57 & 0.50 & - & - \\
\cmidrule{2-10}
~ & UMX & S & 5.7 & 2.16 & 1.38 & 1.83 & \underline{0.63} & - & - \\
~ & MetricGAN+ & S & 3.6 & \textbf{2.83} & \textbf{2.18} & \textbf{2.61} & 0.60 & - & - \\
\cmidrule{2-10}
~ & RVAE-VEM & U-NA & 5.8 & 2.27 & 1.54 & 1.98 & 0.62 & 300 & 27.91 \\
\cmidrule{2-10}
~ & RVAE-LV & U-NA\,/\,U-ND\,/\,U-NDA & 5.4\,/\,5.3\,/\,\textbf{6.2} & 2.31\,/\,2.25\,/\,\underline{2.38} & 1.53\,/\,1.53\,/\,\underline{1.65} & 2.01\,/\,1.95\,/\,\underline{2.07} & \textbf{0.65}\,/\,0.60\,/\,0.62 & 1000\,/\,0\,/\,190 & 89.42\,/\,\textbf{0.02}\,/\, \underline{17.42} \\
~ & RVAE-NO & U-NA\,/\,U-ND\,/\,U-NDA & \underline{6.0}\,/\,3.7\,/\,5.8 & 2.33\,/\,2.11\,/\,2.31 & 1.56\,/\,1.37\,/\,1.54 & 2.04\,/\,1.81\,/\,2.02 & \textbf{0.65}\,/\,0.58\,/\,\underline{0.63} & 1000\,/\,0\,/\,500 & 89.34\,/\,\textbf{0.02}\,/\, 45.54 \\
~ & RVAE-NOLV & U-NA\,/\,U-ND\,/\,U-NDA & 5.5\,/\,4.9\,/\,\textbf{6.2} & 2.31\,/\,2.11\,/\,2.29 & 1.53\,/\,1.42\,/\,1.56 & 2.01\,/\,1.83\,/\,2.00 & \textbf{0.65}\,/\,0.60\,/\,0.62 & 1000\,/\,0\,/\,500 & 90.98\,/\,\textbf{0.02}\,/\, 45.92 \\
\midrule[1pt]
\multirow{11}*{\rotatebox{90}{VB-DMD}} & Noisy mix. & -  & 8.4 & 3.02 & 1.97 & 2.88 & 0.79 & - & - \\
\cmidrule{2-10}
~ & UMX & S & 14.0 & 3.18 & 2.35 & 3.08 & \underline{0.83} & - & - \\
~ & MetricGAN+ & S & 8.5 & \textbf{3.59} & \textbf{3.13} & \textbf{3.63} & \underline{0.83} & - & - \\
~ & CDiffuSE & S & 12.6 & - & 2.46 & - & 0.79 & - & - \\
~ & SGMSE+ & S & 17.3 & - & \underline{2.93} & - & \textbf{0.87} & - & 3.39 \\
\cmidrule{2-10}
~ & NyTT Xtra & U-ND & \underline{17.7} & - & 2.30 & - & - & - & - \\
~ & MetricGAN-U & U-ND & 8.2 & 3.20 & 2.45 & 3.11 & 0.77 & - & - \\
~ & RVAE-VEM & U-NA & 17.1 & 3.23 & 2.48 & 3.15 & 0.81 & 100 & 9.55 \\
\cmidrule{2-10}
~ & RVAE-LV & U-NA\,/\,U-ND\,/\,U-NDA & 17.5\,/\,17.4\,/\,\textbf{17.8} & 3.23\,/\,3.24\,/\,3.22 & 2.39\,/\,2.40\,/\,2.38 & 3.15\,/\,3.17\,/\,3.14 & 0.82\,/\,0.81\,/\,0.81 & 900\,/\,0\,/\,25 & 81.62\,/\,\textbf{0.02}\,/\, 2.32 \\
~ & RVAE-NO & U-NA\,/\,U-ND\,/\,U-NDA & 17.3\,/\,16.7\,/\,17.2 & \underline{3.25}\,/\,3.03\,/\,3.06 & 2.40\,/\,2.12\,/\,2.18 & \underline{3.18}\,/\,2.89\,/\,2.93 & 0.82\,/\,0.79\,/\,0.80 & 400\,/\,0\,/\,25 & 36.79\,/\,\textbf{0.02}\,/\, \underline{2.13} \\
~ & RVAE-NOLV & U-NA\,/\,U-ND\,/\,U-NDA & 17.5\,/\,16.9\,/\,17.4 & \underline{3.25}\,/\,3.04\,/\,3.17 & 2.40\,/\,2.14\,/\,2.30 & \underline{3.18}\,/\,2.90\,/\,3.07 & 0.82\,/\,0.79\,/\,0.81 & 800\,/\,0\,/\,95 & 73.24\,/\,\textbf{0.02}\,/\, 8.84 \\
\bottomrule[1pt]
\end{tabular}
}
\label{table:speech-enhancement-results}
\end{table*}

\subsection{Implementation details and training settings}
The RVAE architecture closely follows the one used in \cite{9894060}, with the exception of replacing the bidirectional LSTM (BLSTM) layers in both the encoder and decoder with standard LSTM layers, since we use here the causal version of RVAE \cite{9053164}. The latent vector dimension was set to $L = 16$.

The NO noise model is implemented using an LSTM layer that takes as input at time $t$ the past noisy speech vectors $\mathbf{x}_{1:t-1}$, followed by a multi-layer perceptron (MLP) with a tanh activation function, except for the output layer, which is linear, and which provides the noise log-variance vector $\log \mathbf{v}_{\theta_n, t}(\mathbf{p}_{t})$. The architecture of the NOLV and LV noise model are similar to that of the NO noise model, except that the NOLV model uses two LSTM layers, one to encode information from the past noisy speech vectors $\mathbf{x}_{1:t-1}$ and another one to process the past and present latent vectors $\mathbf{z}_{1:t}$, and the LV noise model uses a single BLSTM layer to encode information from the complete latent vector sequence  $\mathbf{z}_{1:T}$. 

For all training processes, we used the Adam optimizer \cite{KingmaB14} with parameters $\beta_1=0.9, \beta_2=0.99, \epsilon=10^{-9}$. For RVAE pre-training and ND training configuration, we decayed the learning rate (from $5 \times 10^{-4}$ to $10^{-8}$) with a cosine annealing scheduler \cite{loshchilov2017sgdr}. The models are trained in maximum 500 epochs and the validation set is used to select the best models. During the RVAE pre-training, we applied linear warm-up to the KL term in \eqref{eq:RVAE-ELBO} during the first 20 epochs \cite{NIPS2016_6ae07dcb}. 

\subsection{Baselines and evaluation metrics}
We compare our method with both supervised and unsupervised speech enhancement baselines. For supervised baselines, we considered Open-Unmix (UMX) \cite{stoter19} and MetricGAN+ \cite{Fu2021MetricGANAI}, which are BLSTM-based methods, and CDiffuSE \cite{9746901} and SGMSE+ \cite{richter2022speech}, which are diffusion-based methods. For unsupervised baselines, we compared to MetricGAN-U \cite{9747180}, NyTT \cite{9616166}, and RVAE-VEM \cite{9053164,9894060}.

As for the speech enhancement performance metrics, we used the scale-invariant signal-to-distortion ratio (SI-SDR) \cite{8683855} in dB, the perceptual evaluation of speech quality (PESQ) score \cite{941023} (in $[-0.5, 4.5]$), and the extended short-time objective intelligibility (ESTOI) score \cite{5713237} (in $[0,1]$). We also evaluated the computational efficiency of the inference (denoising algorithm) for RVAE-VEM, SGMSE+ and the proposed method (in different configurations) using the average real-time factor (RTF), which is the time required to process 1 second of audio.\footnote{All of the RTF values are computed on NVIDIA Quadro RTX 4000 GPU, in a machine with an Intel(R) Xeon(R) W-2145 CPU @ 3.70GHz and averaged on 10 sequences.} 

\subsection{Experimental results}

The speech enhancement results are reported in Table~\ref{table:speech-enhancement-results}. When tested in the NA configuration, our method has a better performance than the unsupervised baselines on both datasets in terms of the SI-SDR. The results are actually comparable to those of the supervised methods (although this can depend on the metrics), even though we have never used pairs of aligned noisy-clean speech data for training. This shows the ability of the proposed model to adapt to the noise characteristics in the NA configuration. More specifically, in the NA configuration, the RVAE-NO model performs slightly better than the other two models on the WSJ0-QUT dataset while the three noise model variants (NO, NOLV, LV) lead to very similar performance on the VB-DMD dataset. However, the RVAE-NO model and RVAE-NOLV model have a slight drop of performance when trained in ND configuration. This may be due to the mismatch between train and test data in this ND configuration (different types of noise being used for training and test). In contrast, the RVAE-LV model reveals very robust when used in the ND configuration. This may be because estimating the noise variance only from the latent vectors, without using the noisy speech vectors, helps to alleviate the training/test data mismatch issue.

As for the computational cost, in the NA configuration, the RTF mainly depends on the number of iterations run on each test sequence. 
It can be seen from Table \ref{table:speech-enhancement-results} that, in general, achieving good performance in the NA configuration is at the price of very high RTF values. In contrast, in the ND configuration, the inference process only requires a single forward pass of the trained model, resulting in a much lower RTF value of 0.02 for all of the proposed model variants. Due to the time-consuming inverse diffusion process, the state-of-the-art supervised baseline SGMSE+ has an RTF value that is much higher than the proposed unsupervised model in the ND configuration, while the two methods have similar performance in terms of SI-SDR (however, the RTF of SGMSE+ remains much lower than the proposed model in the NA configuration).

Finally, after being trained in the ND configuration, the proposed model can further be fine-tuned on each noisy test sequence, just like in the NA configuration. This new `hybrid' mode is referred to as NDA in Table \ref{table:speech-enhancement-results}. We found that on the VB-DMD dataset, after just a few iterations of fine-tuning, the performance of all the three model variants were greatly improved (over the ND configuration). This is also true on WSJ0-QUT, but at the price of more fine-tuning iterations. 

\section{Conclusion}
We presented a new unsupervised speech enhancement model that uses a DDGM for both speech and noise. 
We tested three different dependencies for the noise model (NO, NOLV, LV), as well as three `training/testing' configurations (NA, ND, and ND + noise adaptation). Experiments show that in the NA configuration, our model outperforms several unsupervised baselines (including RVAE+NMF), and competes well with the supervised baselines. In the ND configuration, our model provides a very fast inference process with only minimal performance degradation (especially for RVAE-LV). Furthermore, the ND + noise adaptation configuration  enables the model to adapt to specific noise types and further improve performance, with much less iterations than in the NA configuration.

\bibliographystyle{IEEEtran}
\bibliography{mybib}

\begin{thebibliography}{10}
\providecommand{\url}[1]{#1}
\csname url@samestyle\endcsname
\providecommand{\newblock}{\relax}
\providecommand{\bibinfo}[2]{#2}
\providecommand{\BIBentrySTDinterwordspacing}{\spaceskip=0pt\relax}
\providecommand{\BIBentryALTinterwordstretchfactor}{4}
\providecommand{\BIBentryALTinterwordspacing}{\spaceskip=\fontdimen2\font plus
\BIBentryALTinterwordstretchfactor\fontdimen3\font minus
  \fontdimen4\font\relax}
\providecommand{\BIBforeignlanguage}[2]{{%
\expandafter\ifx\csname l@#1\endcsname\relax
\typeout{** WARNING: IEEEtran.bst: No hyphenation pattern has been}%
\typeout{** loaded for the language `#1'. Using the pattern for}%
\typeout{** the default language instead.}%
\else
\language=\csname l@#1\endcsname
\fi
#2}}
\providecommand{\BIBdecl}{\relax}
\BIBdecl

\bibitem{speechenhancement}
J.~Benesty, S.~Makino, and J.~Chen, \emph{Speech Enhancement}.\hskip 1em plus
  0.5em minus 0.4em\relax Springer Berlin, Heidelberg, 2005.

\bibitem{loizou2007speech}
P.~C. Loizou, \emph{Speech enhancement: {T}heory and practice}.\hskip 1em plus
  0.5em minus 0.4em\relax CRC press, 2007.

\bibitem{8369155}
D.~Wang and J.~Chen, ``Supervised speech separation based on deep learning: An
  overview,'' \emph{IEEE/ACM Trans. Audio, Speech, Lang. Process.}, vol.~26,
  no.~10, pp. 1702--1726, 2018.

\bibitem{Pascual2017SEGANSE}
S.~Pascual, A.~Bonafonte, and J.~Serr{\`a}, ``{SEGAN}: Speech enhancement
  generative adversarial network,'' in \emph{Proc. Interspeech}, Stockholm,
  Sweden, 2017.

\bibitem{fu2019metricGAN}
S.-W. Fu, C.-F. Liao, Y.~Tsao, and S.-D. Lin, ``Metric{GAN}: Generative
  adversarial networks based black-box metric scores optimization for speech
  enhancement,'' in \emph{Proc. ICML}, Long Beach, CA, 2019.

\bibitem{Fu2021MetricGANAI}
S.-W. Fu, C.~Yu, T.-A. Hsieh, P.~W.~V. Plantinga, M.~Ravanelli, X.~Lu, and
  Y.~Tsao, ``Metric{GAN}+: An improved version of {MetricGAN} for speech
  enhancement,'' in \emph{Proc. Interspeech}, Brno, Czech Republic, 2021.

\bibitem{9746901}
Y.-J. Lu, Z.-Q. Wang, S.~Watanabe, A.~Richard, C.~Yu, and Y.~Tsao,
  ``Conditional diffusion probabilistic model for speech enhancement,'' in
  \emph{Proc. ICASSP}, Singapore, 2022.

\bibitem{richter2022speech}
J.~Richter, S.~Welker, J.-M. Lemercier, B.~Lay, and T.~Gerkmann, ``Speech
  enhancement and dereverberation with diffusion-based generative models,''
  \emph{arXiv preprint arXiv:2208.05830}, 2022.

\bibitem{9104020}
Y.~Xiang and C.~Bao, ``A parallel-data-free speech enhancement method using
  multi-objective learning cycle-consistent generative adversarial network,''
  \emph{IEEE/ACM Trans. Audio, Speech, Lang. Process.}, vol.~28, pp.
  1826--1838, 2020.

\bibitem{9689669}
G.~Yu, Y.~Wang, C.~Zheng, H.~Wang, and Q.~Zhang, ``Cycle{GAN}-based
  non-parallel speech enhancement with an adaptive attention-in-attention
  mechanism,'' in \emph{Proc. APSIPA}, Tokyo, Japan, 2021.

\bibitem{8461530}
Y.~Bando, M.~Mimura, K.~Itoyama, K.~Yoshii, and T.~Kawahara, ``Statistical
  speech enhancement based on probabilistic integration of variational
  autoencoder and non-negative matrix factorization,'' in \emph{Proc. ICASSP},
  Calgary, Canada, 2018.

\bibitem{8516711}
S.~Leglaive, L.~Girin, and R.~Horaud, ``A variance modeling framework based on
  variational autoencoders for speech enhancement,'' in \emph{Proc. MLSP},
  Aalborg, Denmark, 2018.

\bibitem{9894060}
X.~Bie, S.~Leglaive, X.~Alameda-Pineda, and L.~Girin, ``Unsupervised speech
  enhancement using dynamical variational autoencoders,'' \emph{IEEE/ACM Trans.
  Audio, Speech, Lang. Process.}, vol.~30, pp. 2993--3007, 2022.

\bibitem{Alamdari2019ImprovingDS}
N.~Alamdari, A.~Azarang, and N.~Kehtarnavaz, ``Improving deep speech denoising
  by noisy2noisy signal mapping,'' \emph{Applied Acoustics}, vol. 172, p.
  107631, 2019.

\bibitem{Kashyap2021SpeechDW}
M.~M. Kashyap, A.~Tambwekar, K.~Manohara, and S.~Natarajan, ``Speech denoising
  without clean training data: a noise2noise approach,'' in \emph{Proc.
  Interspeech}, Brno, Czech Republic, 2021.

\bibitem{9616166}
T.~Fujimura, Y.~Koizumi, K.~Yatabe, and R.~Miyazaki, ``Noisy-target training: A
  training strategy for {DNN}-based speech enhancement without clean speech,''
  in \emph{Proc. EUSIPCO}, virtual conf., 2021.

\bibitem{9747180}
S.-W. Fu, C.~Yu, K.-H. Hung, M.~Ravanelli, and Y.~Tsao, ``Metric{GAN-U}:
  Unsupervised speech enhancement / dereverberation based only on noisy /
  reverberated speech,'' in \emph{Proc. ICASSP}, Singapore, 2022.

\bibitem{Pariente2019ASP}
M.~Pariente, A.~Deleforge, and E.~Vincent, ``A statistically principled and
  computationally efficient approach to speech enhancement using variational
  autoencoders,'' in \emph{Proc. Interspeech}, Graz, Austria, 2019.

\bibitem{10095237}
M.~Sadeghi and R.~Serizel, ``Fast and efficient speech enhancement with
  variational autoencoders,'' in \emph{Proc. ICASSP}, 2023, pp. 1--5.

\bibitem{8682546}
S.~Leglaive, U.~Şimşekli, A.~Liutkus, L.~Girin, and R.~Horaud, ``Speech
  enhancement with variational autoencoders and alpha-stable distributions,''
  in \emph{Proc. ICASSP}, Brighton, UK, 2019.

\bibitem{9414363}
G.~Carbajal, J.~Richter, and T.~Gerkmann, ``Guided variational autoencoder for
  speech enhancement with a supervised classifier,'' in \emph{Proc. ICASSP},
  Toronto, Canada, 2021.

\bibitem{8683704}
S.~Leglaive, L.~Girin, and R.~Horaud, ``Semi-supervised multichannel speech
  enhancement with variational autoencoders and non-negative matrix
  factorization,'' in \emph{Proc. ICASSP}, Brighton, UK, 2019.

\bibitem{9747036}
Y.~Xiang, J.~L. Højvang, M.~H. Rasmussen, and M.~G. Christensen, ``A bayesian
  permutation training deep representation learning method for speech
  enhancement with variational autoencoder,'' in \emph{Proc. ICASSP}, 2022, pp.
  381--385.

\bibitem{MAL-089}
L.~Girin, S.~Leglaive, X.~Bie, J.~Diard, T.~Hueber, and X.~Alameda-Pineda,
  ``Dynamical variational autoencoders: A comprehensive review,'' \emph{Found.
  Trends Mach. Learn.}, vol.~15, no. 1-2, pp. 1--175, 2021.

\bibitem{Bie2021ABO}
X.~Bie, L.~Girin, S.~Leglaive, T.~Hueber, and X.~Alameda-Pineda, ``A benchmark
  of dynamical variational autoencoders applied to speech spectrogram
  modeling,'' in \emph{Proc. Interspeech}, Brno, Czech Republic, 2021.

\bibitem{9053164}
S.~Leglaive, X.~Alameda-Pineda, L.~Girin, and R.~Horaud, ``A recurrent
  variational autoencoder for speech enhancement,'' in \emph{Proc. ICASSP},
  Barcelona, Spain, 2020.

\bibitem{6797100}
C.~Févotte, N.~Bertin, and J.-L. Durrieu, ``Nonnegative matrix factorization
  with the {I}takura-{S}aito divergence: With application to music analysis,''
  \emph{Neural Comp.}, vol.~21, no.~3, pp. 793--830, 2009.

\bibitem{5720325}
A.~Liutkus, R.~Badeau, and G.~Richard, ``Gaussian processes for underdetermined
  source separation,'' \emph{IEEE Trans. Signal Process.}, vol.~59, no.~7, pp.
  3155--3167, 2011.

\bibitem{Valentini-Botinhao+2016}
C.~Valentini-Botinhao, X.~Wang, S.~Takaki, and J.~Yamagishi, ``Investigating
  {RNN}-based speech enhancement methods for noise-robust text-to-speech,'' in
  \emph{Proc. SSW}, Sunnyvale, CA, 2016.

\bibitem{WSJ0}
\BIBentryALTinterwordspacing
J.~S. Garofolo, D.~Graff, D.~Paul, and D.~Pallett, ``{CSR-I (WSJ0) Sennheiser
  LDC93S6B},'' \emph{Philadelphia: Linguistic Data Consortium}, 1993. [Online].
  Available: \url{https://catalog.ldc.upenn.edu/LDC93S6B}
\BIBentrySTDinterwordspacing

\bibitem{Dean2015TheQP}
D.~Dean, A.~Kanagasundaram, H.~Ghaemmaghami, M.~H. Rahman, and S.~Sridharan,
  ``The {QUT-NOISE-SRE} protocol for the evaluation of noisy speaker
  recognition,'' in \emph{Proc. Interspeech}, Dresden, Germany, 2015.

\bibitem{Veaux2013TheVB}
C.~Veaux, J.~Yamagishi, and S.~King, ``The {V}oice {B}ank corpus: Design,
  collection and data analysis of a large regional accent speech database,'' in
  \emph{Proc. COCOSDA}, Gurgaon, India, 2013.

\bibitem{Thiemann2013TheDE}
J.~Thiemann, N.~Ito, and E.~Vincent, ``The diverse environments multi-channel
  acoustic noise database ({DEMAND}): A database of multichannel environmental
  noise recordings,'' \emph{J. Acoust. Soc. Am.}, vol. 133, pp. 3591--3591,
  2013.

\bibitem{KingmaB14}
D.~P. Kingma and J.~Ba, ``Adam: A method for stochastic optimization,'' in
  \emph{Proc. ICLR}, San Diego, CA, 2015.

\bibitem{loshchilov2017sgdr}
I.~Loshchilov and F.~Hutter, ``{SGDR}: Stochastic gradient descent with warm
  restarts,'' in \emph{Proc. ICLR}, Toulon, France, 2017.

\bibitem{NIPS2016_6ae07dcb}
C.~K. S{\o}nderby, T.~Raiko, L.~Maal{\o}e, S.~K. S{\o}nderby, and O.~Winther,
  ``Ladder variational autoencoders,'' in \emph{Proc. NIPS}, Barcelona, Spain,
  2016.

\bibitem{stoter19}
F.-R. St{\"o}ter, S.~Uhlich, A.~Liutkus, and Y.~Mitsufuji, ``{Open-Unmix} --
  {A} reference implementation for music source separation,'' \emph{J. Open
  Source Software}, vol.~4, no.~41, p. 1667, 2019.

\bibitem{8683855}
J.~L. Roux, S.~Wisdom, H.~Erdogan, and J.~R. Hershey, ``{SDR} – {H}alf-baked
  or well done?'' in \emph{Proc. ICASSP}, Brighton, UK, 2019.

\bibitem{941023}
A.~Rix, J.~Beerends, M.~Hollier, and A.~Hekstra, ``Perceptual evaluation of
  speech quality ({PESQ}) -- {A} new method for speech quality assessment of
  telephone networks and codecs,'' in \emph{Proc. ICASSP}, Salt Lake City, UT,
  2001.

\bibitem{5713237}
C.~Taal, R.~Hendriks, R.~Heusdens, and J.~Jensen, ``An algorithm for
  intelligibility prediction of time–frequency weighted noisy speech,''
  \emph{IEEE Trans. Audio, Speech, Lang. Process.}, vol.~19, no.~7, pp.
  2125--2136, 2011.

\end{thebibliography}

\end{document}